\documentclass[journal=jacsat,manuscript=article]{achemso}
\usepackage[version=3]{mhchem} 
\usepackage{bibentry}

\author{Kui Gong}
\affiliation{Centre for the Physics of Materials and
Department of Physics, McGill University, Montreal, PQ, Canada, H3A 2T8}
\alsoaffiliation{School of Materials Science and Engineering, University of Science and Technology Beijing,
Beijing, 100083 China}

\author{Lei Zhang}
\email{zhanglei@physics.mcgill.ca}
\affiliation{Centre for the Physics of Materials and
Department of Physics, McGill University, Montreal, PQ, Canada, H3A 2T8}

\author{Wei Ji}
\affiliation{Department of Physics and Beijing Key Laboratory of Optoelectronic Functional Materials and Micro-nano Devices,
Renmin University of China, Beijing 100872, China}

\author{Hong Guo}
\affiliation{Centre for the Physics of Materials and
Department of Physics, McGill University, Montreal, PQ, Canada, H3A 2T8}

\title{Electrical contacts to monolayer black Phosphorus: a first principles investigation}


\begin{document}

\begin{abstract}
We report first principles theoretical investigations of possible metal contacts to monolayer black phosphorus (BP). By analyzing lattice geometry, five metal surfaces are found to have minimal lattice mismatch with BP: Cu(111), Zn(0001), In(110), Ta(110) and Nb(110). Further studies indicate Ta and Nb bond strongly with monolayer BP causing substantial bond distortions, but the combined Ta-BP and Nb-BP form good metal surfaces to contact a second layer BP. By analyzing the geometry, bonding, electronic structure, charge transfer, potential and band bending, it is concluded that Cu(111) is the best candidate to form excellent Ohmic contact to monolayer BP. Other four metal surfaces or combined surfaces also provide viable structures to form metal/BP contacts, but they have Schottky character.
\end{abstract}

Two dimensional (2D) layered materials such as graphene and transition metal dichalcogenides (TMDC)
have attracted great attention\cite{graphene,Splendiani,Mak1,ChemicalReview,PRX} as emerging device materials for nanoelectronics due to their novel mechanical, electrical and optical properties. Most recently, layered black phosphorus (BP) - a new and apparently stable elementary 2D material, has been successfully fabricated experimentaly\cite{FuDan,bulk-phosphorus,singapore,Netherlands,Han-Liu}. BP is an allotrope of phosphorus and can be mechanically exfoliated since the layers of BP are held together by van der Waals (vdW) interaction. Different from graphene, the atoms in a single layer BP are not sitting in a flat-land: instead they form a buckled hexagonal structure by covalence bonds\cite{bulk-phosphorus}, as shown in Figure 1(a). Few-layer BP has been predicted as an ideal direct band gap material at the $\Gamma$ point\cite{Wei-Ji}, a property that is very important for electronic and optical applications. Few-layer BP field effect transistors (FET) having high mobility at around 1000 cm$^2$V$^{-1}$s$^{-1}$, has been reported experimentally.\cite{FuDan,singapore,Netherlands,Han-Liu}. However, a large Schottky barrier for n-doped multi-layer BP FET was found to seriously and detrimentally affect the current-voltage characteristics at small bias\cite{FuDan} in the experimental device. Indeed, in both the traditional microelectronics as well as the emerging nanoelectronics, designing proper metal-semiconductor contacts is a crucial problem of device physics\cite{2006contact,2008contactNW,Schottky-barrier}. A large potential barrier or Schottky barrier at the metal-semiconductor contact has a significant negative influence on charge transport in FETs.

For emerging nano-materials such as carbon nanotubes\cite{2005nanotube,2006Nanotube}, graphene \cite{Kelly1,Kelly2,Kelly3} and TMDC\cite{Igor,indium,zhang-zhengyu,2014contactMoS2,MoS2-Sc}, tremendous theoretical and experimental efforts have been devoted to understand the metal contacts. On the other hand, the metal-BP contacts have not received systematic investigation so far, and it is purpose of this work to fill the void. In particular, we aim to determine metal-BP contacts that are Ohmic in order for BP to realize its full potential as a new and emerging electronics material.

Using the density functional theory (DFT) total energy approach, we have systematically investigated atomic structures of  monolayer BP on several important metal substrates including Ta(110), Nb(110), Cu(111), Zn(0001), and In(110).  These metal surfaces cover a substantial range of work functions and have relatively minimal lattice mismatch with BP, therefore one expects them to serve as possible contact material for BP. We find that monolayer BP are stable on pristine Cu, Zn, and In substrates. For Ta and Nb, the strong interaction between BP and Ta/Nb induces substantial P-P bond distortions in the BP, as such these metals do not make good contacts to monolayer BP. On the other hand, we found that a bi-layer BP makes excellent contact to Ta/Nb substrates. For the materials we investigated, it is predicted that the Cu/BP contact forms almost perfect Ohmic character.

\begin{figure}[t!]
\centering
\includegraphics[width=8.5cm,height=6.0cm]{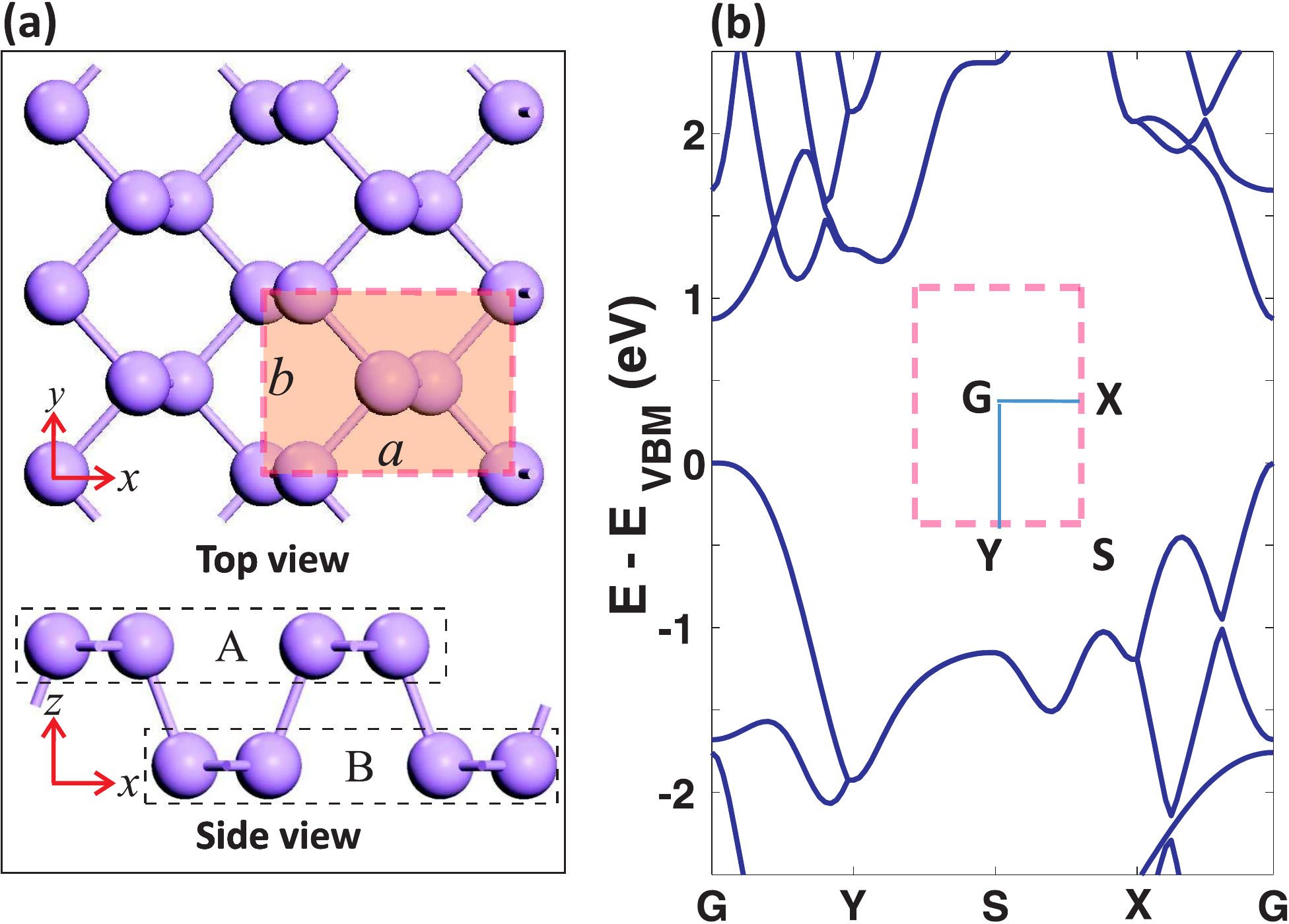}
\caption{(Color online) (a) Top and side view of monolayer BP. The shadowed area with four P atoms is a unit cell of monolayer BP. In the side view, atoms in monolayer BP are classified as A/B sub-layers. (b) The calculated band structure of monolayer BP and its Brillouin zone with high symmetry points.} \label{fig1}
\end{figure}

\textbf{{Computation method and structural relaxation of BP.}} Our DFT calculations are carried out with the projector augmented wave (PAW) \cite{PAW1,PAW2} and the optB88-vdW method\cite{optb881,optb882}, where van der Waals interaction was considered at the vdW-DF level with optB88 for the exchange functional, as implemented in the Vienna ab initio simulation package (VASP) \cite{vasp}. The lattice constant of metals are directly taken from experimental values while the lattice constant of free standing monolayer and bilayer BP are obtained by structure relaxation. To investigate the metal/BP contact, we have chosen a supercell that contains a slab of five layers of metal atoms, a few layer of BP sheets absorbed on the metal slab, and a vacuum region of ~$15${\AA} thick. The dipole correction has been included to avoid spurious interactions between periodic images of the slab. For structure relaxation, we fixed the atoms in the two bottom metal layers at their respective bulk position while all other atoms are fully relaxed until the residual force on each atom is smaller than $0.01$eV/{\AA}. A fine k-mesh density at $0.08$/{\AA} and energy cutoff at $500$eV were used to ensure numerical accuracy. The relaxed monolayer BP is shown in Figure 1(a) and we found the lattice constant to be $a$=$4.58${\AA} and $b$=$3.32${\AA}, in agreement with Ref. \citenum{Wei-Ji}. For simplicity of discussion, in the following we shall call the two sub-layers in Fig.1(a) as A and B (see side view). Fig.1(b) plots the Brillouin zone (BZ) and the calculated band structure showing a direct gap of $0.89$eV.

\begin{table}[!hbpt] \renewcommand{\arraystretch}{1.3}
\caption{Supercell representing the super structure of BP matching with metal substrates. $(n\times m)$ denotes that there are $n$ and $m$ BP unit cells along the $x$ and $y$ directions, respectively. The average distance $d_z$ is the equilibrium separation in the $z$ direction between the monolayer BP and the topmost substrate layer after structure relaxation. For simplicity of notation, the inter-layer distance for bulk BP is also denoted as $d_z$ in the table. $d_m$ is the shortest bond length between the monolayer BP atoms and top-most atoms of substrate as shown in Figure 2(b) and (c). E$_b$ is the binding energy per BP unit cell between the monolayer BP and a given substrate. W is work functions of absorbed monolayer BP including the free-standing one (the first column), respectively. $\Delta E_F$ is the Fermi level shift of free-standing BP at the band bending region.}
\begin{tabular}{lccccccc}
\hline
\hline
 & BP & Ta & Cu & Zn & In & TaP & NbP \\
\hline
Supercell &  & ($1\times1$) & ($1\times3$) & ($1\times4$) & ($1\times3$) & ($1\times1$) & ($1\times1$) \\
\hline
d$_z$ ({\AA})      & 3.20 & 1.76 & 2.31 & 2.87 & 3.00 & 3.04 & 3.04 \\
d$_m$ ({\AA})      &      & 2.44 & 2.36 & 2.76 & 3.11 & 3.55 & 3.19 \\
E$_b$ (eV)         &      & 4.89 & 1.30 & 0.64 & 0.52 & 0.50 & 0.50 \\
W (eV)             & 4.72 &      & 4.68 & 4.51 & 3.93 & 4.34 & 4.41 \\
$\Delta$E$_F$ (eV) &      &      & -0.04 & -0.21 & -0.79 & -0.38 & -0.31 \\
\hline
\hline
\end{tabular}
\label{table:nonlin}
\end{table}

\textbf{{Structures of Metal/BP contacts.}} Having determined the structure of free standing monolayer BP, we now calculate the structure of the metal/BP contact as schematically shown in Fig.2(a). As a first design rule of the metal/BP contact, we consider metals having a lattice plane that can match the BP structure because a large mismatch induces strain to distort BP. As monolayer BP is not a perfect honeycomb structure, only a few metals were found to nicely lattice match BP. Next, we note that the puckered structure of BP has smaller elastic modulus along the $x$ direction\cite{Wei-Ji} [see Fig.1(a)], indicating that structure deformation can easily occur along $x$ which is not good for our purpose. Indeed, we found that monolayer BP adsorbed on Al(111) and Sc(0001) surfaces have three and four BP unit cells per supercell along the $x$ direction, and the relaxed metal/BP structure has significant P-P bond distortions. Therefore, a second design rule is to search metal/BP structures that have one or two BP unit cells per supercell along the $x$ direction. After going through the periodic table, we choose metals Ta(110), Nb(110), Cu(111), Zn(0001), and In(110) as candidates for the metal/BP contacts which cover a wide range of work functions and have relatively minimal lattice mismatch with BP.

\begin{figure}[t!]
\centering
\includegraphics[width=8.0cm,height=7.0cm,clip=true]{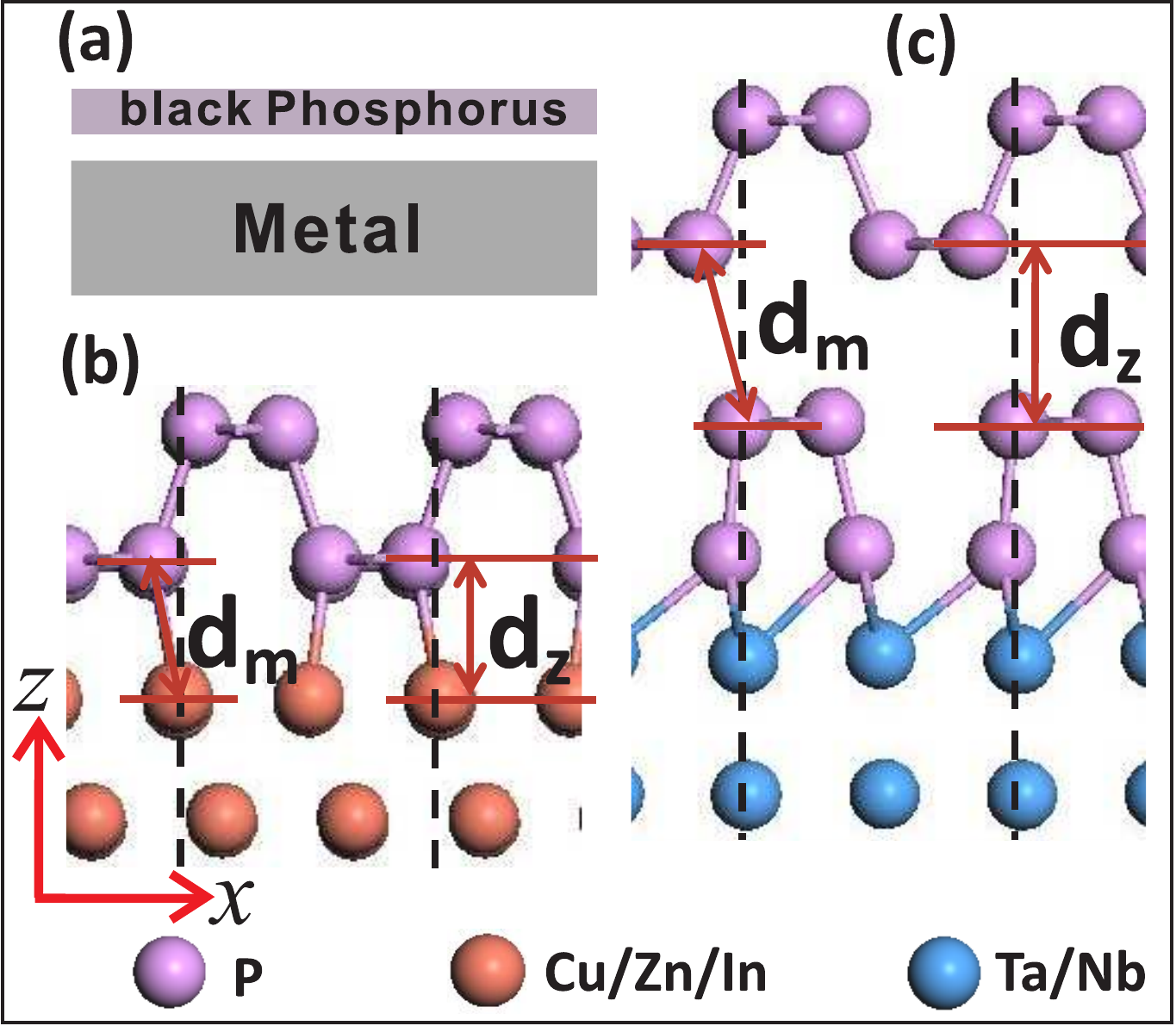}
\caption{(Color online)(a). Schematic plot of the monolayer BP and metal contact region. (b) Side view of monolayer BP absorbed on Cu(111), Zn(0001) and In(110) metal surfaces at the contact region. (c) Side view of monolayer BP absorbed on TaP or NbP electrode at the contact region. The dash line in panel (b) and (c) indicates the supercell in the xz plane.} \label{fig2}
\end{figure}

For each possible candidate of metal/BP contacts, there are a total of seven different initial configurations to be investigated: three for (111) and two for either (110) or (0001) surfaces. From these initial configurations, the most stable metal/BP structures are found by DFT total energy relaxation, shown in Fig. 2(b,c). We find that monolayer BP can only perfectly match with In(110), Cu(111) and Zn(0001) surfaces without significant P-P bond distortion, in the form of (1$\times$3), (1$\times$3), and (1$\times$4) unit cells respectively, as shown in Fig. 2(b). For Ta(110) (1$\times$1) and Nb(110) (1$\times$1) surfaces, it turns out the interaction between P atoms and Ta/Nb atoms are very strong and, as a result, the absorbed monolayer BP is no longer intact due to the presence of broken P-P bonds. Nevertheless, because these two metal surfaces have very small mismatch with the BP atomic structure, i.e (1$\times$1) cell, it is still possible to design good metal/BP contacts. Namely and as shown in Fig.2(c), when a bi-layer BP is in contact with Ta or Nb surfaces, the lower BP layer plus the metal form a \emph{combined substrate} to contact the upper BP layer. The combined substrates are denoted by TaP and NbP in the rest of this paper. We found that a monolayer BP forms very good contact to TaP and NbP.

The calculated equilibrium bonding lengths, binding energies and work functions for all the metal/BP contacts are summarized in \ref{table:nonlin}. The average distance between the substrate and the BP layer along z-axis is $d_z$; the shortest bond between the P atoms and the substrate atoms is $d_m$ (see Fig.2). We characterize the contact strength using the metal/BP binding energy per BP unit cell defined by $E_b = (E_{BP} + E_{sub} - E_{BP-sub})/n$, where $E_{BP-sub}$ is the total energy of BP absorbed on substrate, $E_{BP}, E_{sub}$ are the total energies of BP and substrate, respectively (obtained from the relaxed structure). Here $n$ gives out the number of BP unit cells in the super structure. Among the pure metal/BP contacts (Cu, Zn, In), Cu/BP has the smallest $d_z=2.31${\AA} and $d_m=2.36${\AA}, and the largest binding energy $E_b=1.30$eV. Three other contacts, In/BP, TaP/BP and NbP/BP,  have almost the same $d_z$ and $E_b$ suggesting that they have similar contact properties. For Ta/BP, $E_b=4.89$eV which is much larger than other metal/BP contacts and its $d_z=1.7${\AA} which is much smaller than others: this is because each P atom is bonded to two or more Ta atoms in Ta/BP. Similar situation is found for the Nb substrate (not listed in \ref{table:nonlin}). Such a strong bonding in Ta/BP and Nb/BP induces P-P distortions, making them unsuitable for metal/BP contacts. On the other hand, for the combined contacts TaP/BP and NbP/BP, $d_z=3.04${\AA} which is large enough such that P-P bonds in the upper BP (see Fig.2c) are not broken and, at the same time, this value is smaller than $d_z=3.20${\AA} of the bulk BP, strongly indicate that the combined substrates can make good metal/BP contacts. In the follow we quantify properties of the metal/BP contacts using partial density of states (PDOS), charge density and electronic potential using Cu, Zn, TaP substrates as other contacts have similar properties.

\begin{figure}[t!]
\centering
\includegraphics[width=8.5cm,height=6cm]{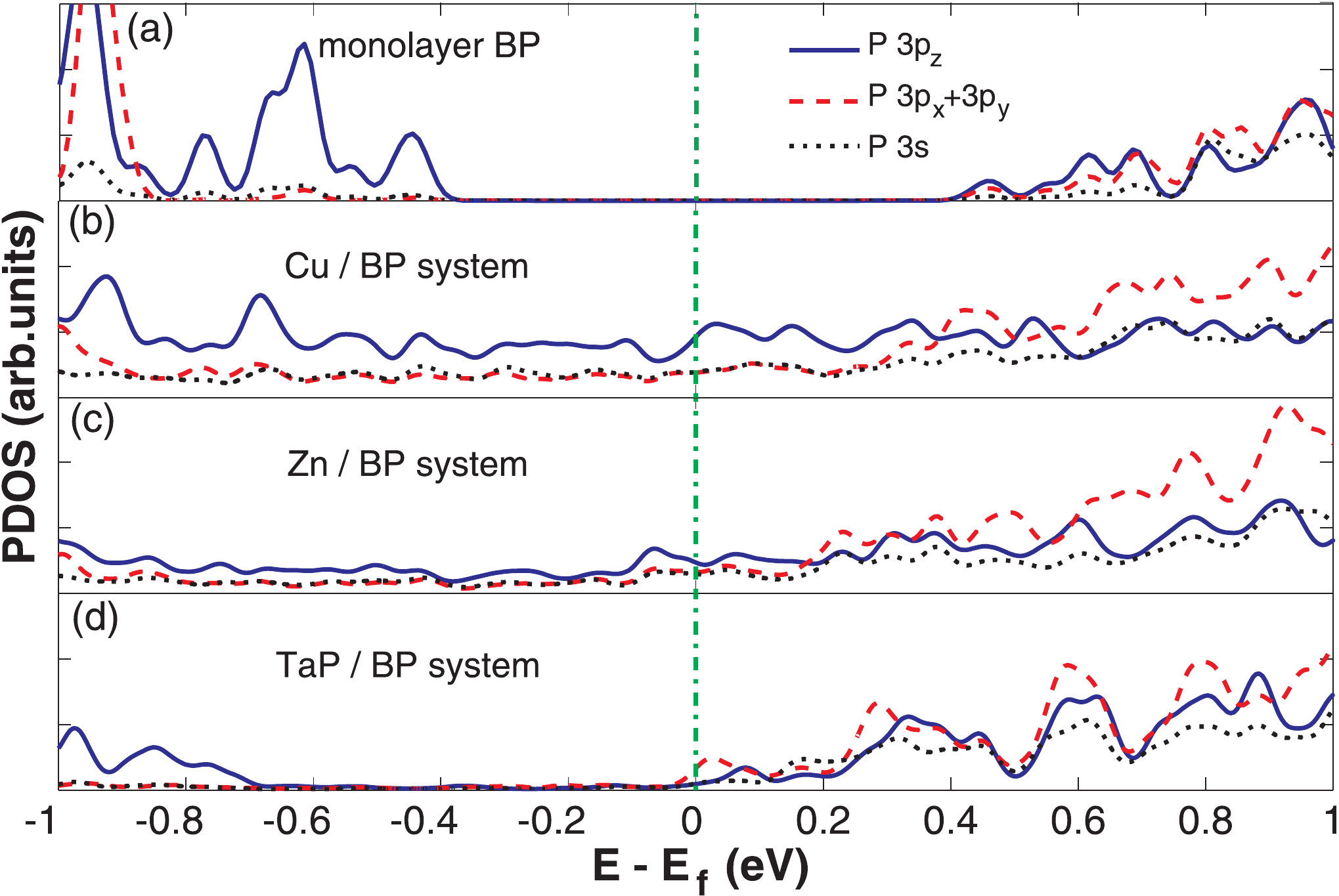}
\caption{(Color online) Partial density of states (PDOS) of P atoms, in (a) the monolayer BP, (b) the Cu/BP system, (c) the Zn/BP system, and (d) the TaP/BP system (The P atoms in the top layer). The blue solid line, red dash line and black dot line represent 3p$_z$ orbital, 3p$_x$+3p$_y$ orbital, and 3s orbital of top layer P atoms as indicated by the legend in (a). The vertical green dash line indicates the Fermi level.} \label{fig3}
\end{figure}

\textbf{{Electronic structure of metal/BP contacts.}} Having obtained the atomic structures of the metal/BP contacts, we now investigate their electronic properties. Figure 3 plots the calculated PDOS projected on selected orbitals of the P atoms for monolayer BP, Cu/BP, Zn/BP and TaP/BP. The PDOS of free standing monolayer BP (Fig.3a) has a gap around the Fermi level - consistent with its band structure (Fig.1b), the valence state is dominated by the $3p_z$ state while the bottom conduction PDOS is contributed by all types of $3p$ orbital. The PDOS of Cu/BP, Zn/BP and TaP/BP has a metallic character as shown in Fig.3(b,c,d). From observing the peak of $3p_x+3P_y$ PDOS, we know the Fermi level of Cu/BP and Zn/BP moves upward by about $0.1$eV and $0.2$eV with respect to the bottom conduction band of the free standing BP, respectively. The PDOS of Cu/BP (Fig.3b) near the Fermi level is much lager than that of Zn/BP (Fig.3c) due to a significant increase of contribution from the $3p_z$ states. Since the Cu-BP bonding distance ($d_z$, see Fig.2b and Table-1) is significantly smaller than that of Zn-BP, the $3p_z$ states of P atoms have much stronger interaction with the $3d$ states of Cu than with Zn. As a result, Cu/BP is a better contact than Zn/BP since good device contacts should maximize overlap between states at either sides of the contact interface. As for the combined contact TaP/BP (Fig.3d), the in-plane states $3p_x, 3p_y$ play a more important role than $3p_z$ state in comparison to Cu and Zn. This is because of Bernal stacking of the top BP layer and the TaP\cite{Wei-Ji}, as a result the overlap of in-plane $3p_{x/y}$ between the two BP layers are much stronger than that of $3p_z$ states.

\begin{figure}
\centering
\includegraphics[width=8.5cm,height=9.8cm]{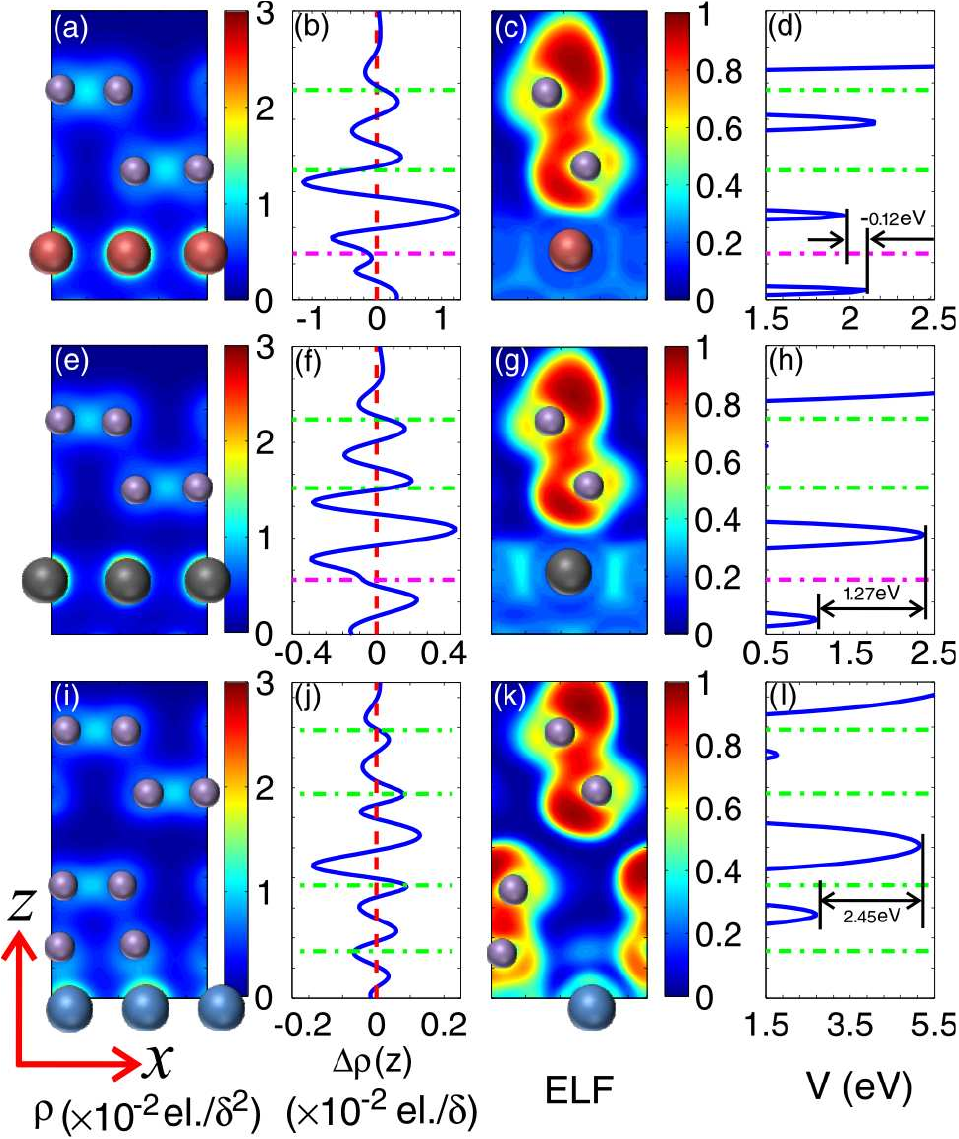}
\caption{(Color online) Electronic structure at the interface of the contact region, (a)-(d) for Cu substrate, (e)-(h) for Zn substrate, and (i)-(l) for TaP substrate. The left side contour plots represent charge density $\rho$ which is integrated in y-direction and normalized to one BP unit cell. The unit is number of electron per area (denoted as $el./\delta^2$) The middle panels (b), (f), and (j) are charge density difference $\Delta\rho$(z), which is integrated over the x- and y-direction and normalized to one BP unit cell. The middle panels (c), (g), and (k) are contour plots of electron localization function (ELF) in a slice along the y-axis that crosses both P atoms and metal atoms. The panels (d), (h), and (l) are average local potential $V$. The average potential barriers at the BP and metal interface are also indicated. $\delta$ is the length of real space grid, equals to 0.065{\AA}. The green dashed lines represent the position of each average P atoms layer and the pink dashed lines represent the position of average metal surface after structure relaxation.} \label{fig4}
\end{figure}
The interaction between BP and metal is quantified by calculating the charge density. To compare different contacts, we integrate the density along y-direction (see Fig.1a) and normalize it per unit cell of BP. The corresponding results for Cu/BP and Zn/BP are presented in the left panels of Fig. 4(a,e), respectively. Here we clearly observe that the normalized charge density $\rho$ at the interface of Cu/BP is much larger than that of Zn/BP interface (by the color coding). As usual, interface dipoles are formed due to interaction of charge carriers at the interface which is visualized from the electron rearrangement defined as $\Delta\rho(z)= (\rho_{sub-BP}$(z) - $\rho_{sub}(z) - \rho_{BP}(z))/n$, where $\rho_{sub}(z)$ and $\rho_{BP}(z)$ are densities of substrate and monolayer BP. Here, negative (positive) $\Delta\rho$(z) stands for charge depletion (accumulation) in the x-y plane, and plotted in Fig.4(b,f). The large net charge accumulation between B-sub-layer of BP (lower P atoms in BP, see Fig.1a) and topmost layer of metal suggests that covalence bonds have been formed. In addition, the magnitude of $\Delta\rho(z)$ in Cu/BP is twice as large than that of Zn/BP, indicating that the strength of covalence bond of Cu/BP is significantly larger than that in Zn/BP.

A vivid physical picture of the chemical bonds at the metal/BP interface can be established by investigating the electron localization function (ELF) for the slice that crosses both P atoms and metal atoms in the y-direction, shown in Fig.4(c,g,k). Physically, ELF measures the extent of spacial localization of a reference electron, with upper limit $ELF({\bf r})=1$ corresponding to perfect localization (lone pairs) and $ELF({\bf r})=1/2$ corresponding to electron-gas-like pair\cite{ELF}. Clearly, from Fig.4(c) the upper P atom (A-layer in Fig.1a) has significantly larger ELF than the lower P atom (B-layer in Fig.1a), this is reasonable because the lower P atom is closer to Cu contact, which make their electrons more delocalized. Similar conclusion is true for the Zn/BP contact, but the ELF of Zn/BP is larger than that of Cu/BP - again indicating a weaker covalent bond between Zn-BP than Cu-BP. As for the combined contact TaP/BP whose ELF is shown in Fig.4(k), there is a delocalized region (dark blue) between the top BP layer and TaP, indicating that the chemical bond between the two BP layers to be very weak and the dominating interaction at the TaP/BP interface is largely via the vdW force. This is why the dipoles between the two BP layers in Figure 4(j) is $3-6$ times smaller than that between BP and Zn or Cu.

From the calculated electronic structure, we obtain a most important parameter for charge injection from metal to BP, namely the potential barrier at the metal/BP interface. The electronic potential (ionic and Hartree contributions) averaged over the x-y plane across the lower P atoms of the BP which are in direct contact to the metal (e.g. the B sub-layer, see Fig.1a), is shown in Fig.4$(d,h)$ for Cu/BP and Zn/BP.
For TaP/BP, the potential is averaged over the P atoms in direct contact to the TaP substrate as shown in Fig.4$(l)$.  Since ionic contribution is included, the highest averaged potential appears at the middle of each two neighboring atomic layers. We define a quantity $\Delta V$ to be the difference between highest averaged potential at the contact interface and the highest averaged potential at the metal surface, as indicated by the black arrows. It comes as a pleasant surprise that $\Delta V = -0.12$eV is \emph{negative} for Cu/BP (Fig. 4d), indicating electrons can be easily injected from Cu to the BP without any potential barrier. On the other hand, Zn/BP and TaP/BP all have positive $\Delta V$ at $1.27$eV and $2.45$eV, respectively. We conclude that Cu/BP is an excellent Ohmic contact and the other contacts have large potential barriers but electron injection into BP is still possible due to non-zero PDOS at the Fermi level.

\textbf{{Discussion and summary.}} Having determined several metal/BP contacts with Cu/BP the best, here we qualitatively discuss a possible application of them in a current-in-plane (CIP) device model. We anticipate a free standing monolayer BP to be the channel material that is connected to a metal/BP contact, as plotted Fig.5. Clearly, charge transfer takes place between the metal/BP contact and the free standing BP - band bending occurs as schematically shown. The band bending can be estimated by the Fermi level difference\cite{Kelly2} between the metal/BP contact and the free standing BP: $\Delta E_F = W - W_{BP}$ where $W$, $W_{BP}$ are the work functions of metal/BP and free standing BP, respectively. If $\Delta E_F > 0 $, electrons transfer from the free standing BP to the metal/BP contact and the channel is p-type. When $\Delta E_F < 0$, the BP channel is n-type. As tabulated in Table-1, we found $\Delta E_F<0$ for all five viable metal/BP contacts hence these contacts tend to produce n-type CIP devices without further doping. Importantly, the Cu/BP contact has the smallest band bending, $\Delta E_F = -0.04$eV. Interestingly, TaP/BP and NbP/BP contacts also have rather small bending, $\Delta E_F\approx -0.3\sim-0.4$eV. From the CIP point of view, the smaller the shift the better the contact.

\begin{figure}
\centering
\includegraphics[width=8cm,height=4.5cm]{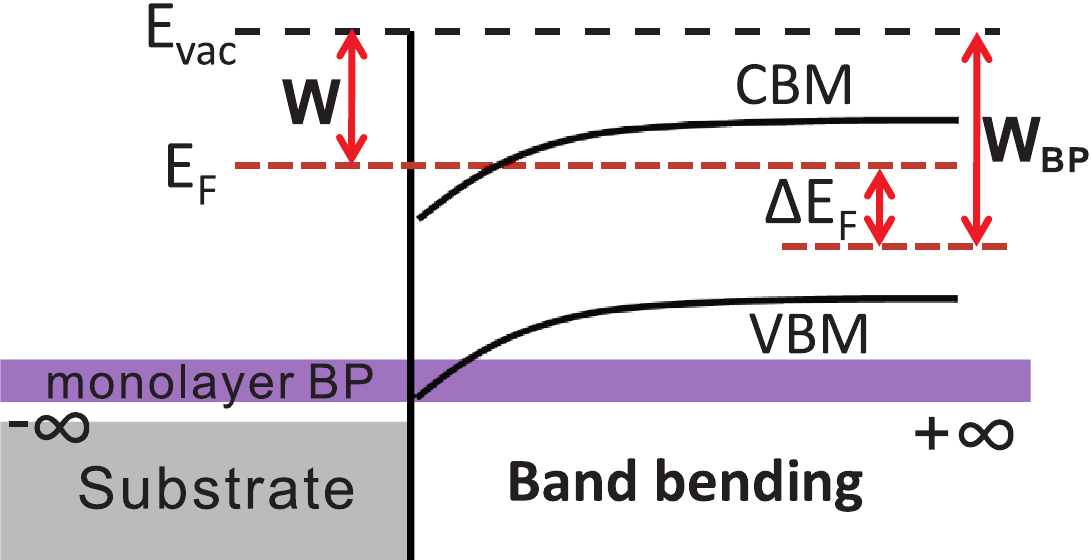}
\caption{(Color online) Schematic plot of a current-in-plane (CIP) device model where the metal/BP contact is connected to a free standing BP channel extending to the right. $W$ and $W_{BP}$ denote work functions of the metal/BP contact and the free standing BP, respectively. For all the metal/BP contacts in Table-1, we found $W< W_{BP}$. $E_{vac}$ denotes the vacuum level, VBM and CBM denote valence band maximum and conduction band minimum, respectively. The black lines qualitatively indicate the band bending.} \label{fig5}
\end{figure}

In summary, by atomistic calculations we have systematically investigated the possible metal/BP contacts which allows us to design viable structures for future device applications. Our design rule starts from selecting metals that have a lattice plane matching the BP structure so that strain at the metal/BP interface is minimized. Of the five metals that satisfy this condition, Ta and Nb are found to form strong bonds with monolayer BP to generate substantial P-P bond distortions in the BP, for this reason we further considered TaP and NbP as combined metal substrates. We predict that monolayer BP absorbed on Cu(111), Zn(0001), In(110), TaP(110) and NbP(110) form viable metal/BP contacts. From the calculated geometry, bonding structure, density of states, charge transfer and potential barriers at the metal/BP interface,we predict that Cu/BP is an excellent Ohmic contact and the rest are Schottky contacts where the electronic potential barrier increases with the averaged distance between monolayer BP and the metal substrate. For the CIP device model where a free standing monolayer BP is the channel material, the estimated band bending property suggests intrinsically n-type device for all the contacts and, in particular, the Cu/BP is the most ideal contact to the free standing BP channel. Therefore we conclude that Cu(111) is the best choice as an electrode metal for monolayer BP electronic device applications. The theoretical predictions of this work should be experimentally testable.

\begin{acknowledgement}
This work is supported by NSERC of Canada and University Grant Council (AoE/P-04/08) of the Government of HKSAR (H.G.), the China Scholarship Council (K.G.) and the National Natural Science Foundation of China (NSFC), Grant No 11274380 (W.J.). We thank CLUMEQ, CalcuQuebec and Compute-Canada for providing computation facilities.
\end{acknowledgement}

\bibliography{BP}

\end{document}